\documentstyle[pra,aps]{revtex}
\input tcilatex
\begin{document}

\title{Reducing spatially correlated noise and decoherence with quantum error
correcting codes}
\author{Lu-Ming Duan\thanks{%
Electronic address: lmduan@ustc.edu.cn} and Guang-Can Guo\thanks{%
Electronic address: gcguo@ustc.edu.cn} \\
%EndAName
Physics Department and Nonlinear Science Center,\\
University of Science and Technology of China,\\
Hefei 230026, People's Republic of China}
\maketitle

\begin{abstract}
\baselineskip 24ptIt is shown that the noise process in quantum computation
can be described by spatially correlated decoherence and dissipation. We
demonstrate that the conventional quantum error correcting codes correcting
for single-qubit errors are applicable for reducing spatially correlated
noise.\\

\end{abstract}

\newpage\ \baselineskip 24pt

Quantum computers hold the promise for solving many hard problems much more
effectively than their classical counterparts [1-2]. However, in practice
the inevitable noise and decoherence process will diminish the advantages of
quantum computation [4]. To overcome the fragility of quantum information,
many kinds of techniques have been discovered to combat noise and
decoherence in quantum computers [5-18]. Among these techniques, quantum
error correction is the most important one [5-12].

In quantum error correction schemes, the input state is encoded into a state
in a larger Hilbert space so that it can be recovered from a certain class
of errors. The error operators are identified from physical consideration.
Conventionally, it is assumed that qubits (quantum bits) are decohered
independently. In this circumstance, after a short time interval, the most
important errors are those described by single-qubit operators [19]. Hence,
quantum error correcting codes (QECCs) need only to correct single-qubit or
few-qubit errors [5-12].

The assumption of independent decoherence plays an important role in quantum
error correction schemes [5,10]. Except independent decoherence, there are
possibly other kinds of decoherence, for example, collective decoherence
[15,16]. Alternate schemes, called quantum error avoiding codes, have been
proposed for reducing collective decoherence [16-18]. Are quantum error
correction techniques applicable for reducing collective decoherence? The
answer is shown to be positive for a special collective dephasing model
[20]. In this paper, we consider the most general type of noise of many
qubits. It is shown that to a good approximation, the general noise process
can be described by spatially correlated decoherence and dissipation, which
includes independent decoherence and collective decoherence as its special
cases. The master equation is derived, and we identify the error operators
for general spatially correlated noise, using the quantum trajectory
approach [21,22]. The error operators for spatially correlated noise are no
longer single-qubit operators. They are either a sum of single-qubit
operators (the quantum jump errors) or a tensor product of them ( the
effective evolution error). However, we demonstrate that the conventional
QECCs correcting for single-qubit errors remain valid for reducing this kind
of spatially correlated noise.

We consider noise and decoherence of $L$ qubits. The $l$ qubit is described
by the Pauli operator $\overrightarrow{\sigma }_l$. The most general noise
process of $L$ qubits can be described by the following interaction
Hamiltonian (setting $\hbar =1$): 
\begin{eqnarray}
H_I\left( t\right) &=&g_1\stackunder{l,\alpha }{\sum }\sigma _l^\alpha
\Gamma _l^\alpha \left( t\right) +g_2\stackunder{l,l^{^{\prime }},l\neq
l^{^{\prime }}}{\sum }\stackunder{\alpha _l,\alpha _{l^{^{\prime }}}}{\sum }%
\sigma _l^{\alpha _l}\otimes \sigma _{l^{^{\prime }}}^{\alpha _{l^{^{\prime
}}}}\Gamma _{ll^{^{\prime }}}^{\alpha _l\alpha _{l^{^{\prime }}}}\left(
t\right)  \nonumber  \label{1} \\
&&  \label{1} \\
&&\ \ \ \ +\cdots +g_L\stackunder{\alpha _1,\alpha _2,\cdots \alpha _L}{\sum 
}\sigma _1^{\alpha _1}\otimes \sigma _2^{\alpha _2}\otimes \cdots \otimes
\sigma _L^{\alpha _L}\Gamma _{12\cdots L}^{\alpha _1\alpha _2\cdots \alpha
_L}\left( t\right) ,  \nonumber
\end{eqnarray}
where $\alpha _l=1,2,3$, and $l=1,2,\cdots ,L$ ($\sigma _l^{1,2,3}$
correspond to $\sigma _l^{x,y,z}$, respectively). All the $\Gamma _l^\alpha
\left( t\right) $, $\Gamma _{ll^{^{\prime }}}^{\alpha _l\alpha _{l^{^{\prime
}}}}\left( t\right) ,$ and $\Gamma _{12\cdots N}^{\alpha _1\alpha _2\cdots
\alpha _N}\left( t\right) $, generally dependent of time, are noise terms,
which may be classical stochastic variables or stochastic quantum operators,
corresponding to classical noise or quantum noise, respectively. The
Hamiltonian (1) includes all possible interaction terms between the qubits
and the noisy environment, and it is very complicate. Fortunately, in
practice this complicate description is not necessary. The coupling
coefficients $g_1$, $g_2$, $\cdots $ and $g_n$ normally satisfy the
condition $g_1\gg g_2\gg \cdots \gg g_L$. Hence, the most important noise
always comes from the first term of the right hand side of Eq. (1), if this
term does not reduce to zero due to some special symmetry. As a good
approximation, we can safely drop off all the high order nonlinearities in
Eq. (1). The interaction Hamiltonian is then simplified to 
\begin{equation}
H_I\left( t\right) =g_1\stackunder{l,\alpha }{\sum }\sigma _l^\alpha \Gamma
_l^\alpha \left( t\right) .  \label{2}
\end{equation}
The explicit expressions for the noise terms $\Gamma _l^\alpha \left(
t\right) $ depend on the concrete physical model of quantum computation.
However, it is reasonable to assume that $\Gamma _l^\alpha \left( t\right) $
satisfy the following conditions: 
\begin{equation}
\left\langle \Gamma _l^\alpha \left( t\right) \right\rangle _{env}=0,
\label{3}
\end{equation}
\begin{equation}
\left\langle \Gamma _l^\alpha \left( t\right) \Gamma _{l^{^{\prime }}}^\beta
\left( t^{^{\prime }}\right) \right\rangle _{env}=f_{ll^{^{\prime
}}}^{\alpha \beta }\left( t-t^{^{\prime }}\right) ,  \label{4}
\end{equation}
where $\left\langle {}\right\rangle _{env}$ denotes average over the
environment. In Eq. (4), all the $f_{ll^{^{\prime }}}^{\alpha \beta }\left(
t-t^{^{\prime }}\right) $ are correlation functions. From the Hermiticity of 
$\Gamma _l^\alpha \left( t\right) $, we have 
\begin{equation}
f_{ll^{^{\prime }}}^{\alpha \beta }\left( t\right) =\left[ f_{l^{^{\prime
}}l}^{\beta \alpha }\left( -t\right) \right] ^{*}.  \label{5}
\end{equation}

The general form of the master equation with the interaction Hamiltonian $%
H_I\left( t\right) $ is expressed as [23] 
\begin{equation}
\frac d{dt}\rho \left( t\right) =-\int_0^\infty d\tau tr_{env}\left\{ \left[
H_I\left( t\right) ,\left[ H_I\left( t-\tau \right) ,\rho \left( t\right)
\otimes \rho _{env}\right] \right] \right\} ,  \label{6}
\end{equation}
where $\rho _{env}$ is the density operator of the environment, and $\rho
\left( t\right) $ denotes the reduced density operator of the qubits in the
interaction picture. Substituting the Hamiltonian (2) into Eq. (6), and
using the conditions (3) and (4), we derive the following master equaiton
for noise and decoherence of $L$ qubits:\break 
\begin{eqnarray}
\frac d{dt}\rho \left( t\right)  &=&-\frac i2\stackunder{l,l^{^{\prime
}},\alpha ,\beta }{\sum }B_{l^{^{\prime }}l}^{\beta \alpha }\left[ \sigma
_{l^{^{\prime }}}^\beta \sigma _l^\alpha ,\rho \left( t\right) \right]  
\nonumber \\
&&  \label{7} \\
&&\ \ \ +\stackunder{l,l^{^{\prime }},\alpha ,\beta }{\sum }A_{l^{^{\prime
}}l}^{\beta \alpha }\left[ \sigma _l^\alpha \rho \left( t\right) \sigma
_{l^{^{\prime }}}^\beta -\frac 12\sigma _{l^{^{\prime }}}^\beta \sigma
_l^\alpha \rho \left( t\right) -\frac 12\rho \left( t\right) \sigma
_{l^{^{\prime }}}^\beta \sigma _l^\alpha \right] ,  \nonumber
\end{eqnarray}
where the coefficients 
\begin{equation}
A_{l^{^{\prime }}l}^{\beta \alpha }=g_1^2\int_{-\infty }^{+\infty
}f_{l^{^{\prime }}l}^{\beta \alpha }\left( \tau \right) d\tau ,  \label{8}
\end{equation}
\begin{equation}
B_{l^{^{\prime }}l}^{\beta \alpha }=-ig_1^2\int_0^\infty \left[
f_{l^{^{\prime }}l}^{\beta \alpha }\left( \tau \right) -f_{l^{^{\prime
}}l}^{\beta \alpha }\left( -\tau \right) \right] d\tau .  \label{9}
\end{equation}
From Eq. (5), it follows that the matrixes $A=\left[ A_{l^{^{\prime
}}l}^{\beta \alpha }\right] $ and $B=\left[ B_{l^{^{\prime }}l}^{\beta
\alpha }\right] $ are Hermitian. The first term of the right hand side of
Eq. (7) represents the environment-induced Lamb phase shift, and the second
term represents decoherence and dissipation of the qubits. If the
coefficients $A_{l^{^{\prime }}l}^{\beta \alpha }$ and $B_{l^{^{\prime
}}l}^{\beta \alpha }$ are directly probational to $\delta _{ll^{^{\prime }}}$%
, Eq. (7) describes independent decoherence (in the terminology of Refs.
[15-19]). In contrast, if the correlation terms $A_{l^{^{\prime }}l}^{\beta
\alpha }$ and $B_{l^{^{\prime }}l}^{\beta \alpha }$ with $l\neq l^{^{\prime
}}$ attain the maximum, Eq. (7) represents collective decoherence [15,16].
In general circumstances, the qubits are subject to spatially correlated
decoherence.

We are interested in the problem that to what extent the noise described by
Eq. (7) can be reduced by the conventional quantum error correction
techniques. To examine the problem, we need first to identify all the
first-order error operators for spatially correlated noise. It is convenient
to use the quantum trajectory approach to attain this goal. The quantum
trajectory approach is a recently-developed numerical simulation method for
solving complicate open quantum systems [21,22]. In this approach, the
evolution of the dissipative system is represented by an ensemble of wave
functions that propagate according to the effective Hamiltonian interrupted
by random quantum jumps [24]. To use the language of quantum trajectories,
we need to re-express the master equation (7) in a diagonal form. The
correlation matrix $A$ is Hermitian, hence it can diagonalized by a unitary
matrix $U=\left[ U_{kl}^{\gamma \alpha }\right] $, i.e., we have 
\begin{equation}
A_{l^{^{\prime }}l}^{\beta \alpha }=\stackunder{k,\gamma }{\sum }%
U_{l^{^{\prime }}k}^{\dagger \beta \gamma }\xi _k^\gamma U_{kl}^{\gamma
\alpha },  \label{10}
\end{equation}
where $\xi _k^\gamma $, with $k=1,2,\cdots ,L$ and $\gamma =1,2,3$, are
eigenvalues of the positive-definite Hermitian matrix $A$, which should be
positive real numbers. Define the operators $s_k^\gamma $ by the equation 
\begin{equation}
s_k^\gamma =\stackunder{l,\alpha }{\sum }U_{kl}^{\gamma \alpha }\sigma
_l^\alpha .  \label{11}
\end{equation}
In general, $s_k^\gamma $ are no longer Hermitian operators. With the
transformations (10) and (11), the master equation (7) is rewritten as 
\begin{equation}
\frac d{dt}\rho \left( t\right) =-iH_{eff}\rho \left( t\right) +i\rho \left(
t\right) H_{eff}^{\dagger }+\stackunder{k,\gamma }{\sum }\xi _k^\gamma
s_k^\gamma \rho \left( t\right) \left( s_k^\gamma \right) ^{\dagger },
\label{12}
\end{equation}
where the non-Hermitian effective Hamiltonian is 
\begin{equation}
H_{eff}=\frac 12\stackunder{l,l^{^{\prime }},\alpha ,\beta }{\sum }%
B_{l^{^{\prime }}l}^{\beta \alpha }\sigma _{l^{^{\prime }}}^\beta \sigma
_l^\alpha -\frac i2\stackunder{k,\gamma }{\sum }\xi _k^\gamma \left(
s_k^\gamma \right) ^{\dagger }s_k^\gamma .  \label{13}
\end{equation}
The first term of the effective Hamiltonian is the Hermitian Lamb phase
shift, which in general cannot be simplified by introducing the operators $%
s_k^\gamma $, since the matrixes $A$ and $B$ do not necessarily commute with
each other. The second term of the right hand side of Eq. (13) is the
non-Hermitian damping Hamiltonian.

Suppose that in a finite time $T_0$, we perform $N$ times error corrections.
In a short time interval $\Delta t=\frac{T_0}N$, we need to find all the
error operators up to the first order of $\Delta t$. In the language of
quantum trajectories, the system evolution described by Eq. (12) is
represented by an ensemble of pure states that evolve according to the
effective Hamiltonian (13), interrupted at random times by quantum jumps $%
s_k^\gamma $. Up to the first order of $\Delta t$, the normalized state
after $\Delta t$ will be either 
\begin{equation}
\left| \Psi _k^\gamma \left( \Delta t\right) \right\rangle =\sqrt{\frac{\xi
_k^\gamma \Delta t}{p_k^\gamma }}s_k^\gamma \left| \Psi \left( 0\right)
\right\rangle  \label{14}
\end{equation}
with probability $p_k^\gamma =\xi _k^\gamma \Delta t\left\langle \Psi \left(
0\right) \right| \left( s_k^\gamma \right) ^{\dagger }s_k^\gamma \left| \Psi
\left( 0\right) \right\rangle $ in case of a jump in decay channel $\left(
k,\gamma \right) $ at a random time in the interval $\Delta t$, or 
\begin{eqnarray}
\left| \Psi _0\left( \Delta t\right) \right\rangle &=&\frac 1{\sqrt{p_0}%
}\exp \left( -iH_{eff}\Delta t\right) \left| \Psi \left( 0\right)
\right\rangle  \nonumber \\
&&  \label{15} \\
\ &\approx &\frac 1{\sqrt{p_0}}\left[ 1-\frac{\Delta t}2\stackunder{k,\gamma 
}{\sum }\xi _k^\gamma \left( s_k^\gamma \right) ^{\dagger }s_k^\gamma -\frac{%
i\Delta t}2\stackunder{l,l^{^{\prime }},\alpha ,\beta }{\sum }B_{l^{^{\prime
}}l}^{\beta \alpha }\sigma _{l^{^{\prime }}}^\beta \sigma _l^\alpha \right]
\left| \Psi \left( 0\right) \right\rangle  \nonumber
\end{eqnarray}
with probability $p_0=1-\stackunder{k,\gamma }{\sum }\xi _k^\gamma \Delta
t\left\langle \Psi \left( 0\right) \right| \left( s_k^\gamma \right)
^{\dagger }s_k^\gamma \left| \Psi \left( 0\right) \right\rangle -o\left(
\Delta t^2\right) $ if no jump occurred. Let $Q_0=\frac 1{\sqrt{p_0}}\left[
1-\frac{\Delta t}2\stackunder{k,\gamma }{\sum }\xi _k^\gamma \left(
s_k^\gamma \right) ^{\dagger }s_k^\gamma -\frac{i\Delta t}2\stackunder{%
l,l^{^{\prime }},\alpha ,\beta }{\sum }B_{l^{^{\prime }}l}^{\beta \alpha
}\sigma _{l^{^{\prime }}}^\beta \sigma _l^\alpha \right] $, $Q_{3\left(
k-1\right) +\gamma }=\sqrt{\frac{\xi _k^\gamma \Delta t}{p_k^\gamma }}%
s_k^\gamma $, and $p_{3\left( k-1\right) +\gamma }=p_k^\gamma $, where $%
k=1,2,\cdots ,L$ and $\gamma =1,2,3$. With this notation, the system state
after a short time $\Delta t$ is then represented by the following density
operator 
\begin{equation}
\rho \left( \Delta t\right) =\stackrel{3L}{\stackunder{n=0}{\sum }}%
p_nQ_n\rho \left( 0\right) Q_n^{\dagger }+o\left( \Delta t^2\right) ,
\label{16}
\end{equation}
where $\rho \left( 0\right) =\left| \Psi \left( 0\right) \right\rangle
\left\langle \Psi \left( 0\right) \right| $. In the above equation, $Q_0$
represents the effective evolution error, and $Q_k$ $\left( k=1,2,\cdots
,3L\right) $ represent the quantum jump errors. All the $Q_n$ $\left(
n=0,1,\cdots ,3L\right) $ make a complete set of the first-order error
operators.

For independent decoherence, the correlation coefficients $A_{l^{^{\prime
}}l}^{\beta \alpha }$ and $B_{l^{^{\prime }}l}^{\beta \alpha }$ are directly
proportional to $\delta _{ll^{^{\prime }}}$. All the first-order errors $Q_n$
then reduce to single-qubit operators. For general spatially correlated
decoherence, the first-order errors are no longer single-qubit operators.
The quantum jump errors $Q_k$ $\left( k=1,2,\cdots ,3L\right) $ are
expressed as sums of single-qubit operators, and more seriously, the
effective evolution error $Q_0$ includes the terms that are tensor products
of single-qubit operators. Hence, it is not clear that this kind of
decoherence can be reduced by the conventional QECCs. In fact, it has been
suggested that to combat the effective evolution error, more involved and
less efficient QECCs need be devised [25]. However, here we show that the
conventional QECCs correcting for single-qubit errors remain applicable for
reducing general spatially correlated decoherence, if the error correction
procedure is repeated frequently enough. This can be demonstrated by the
following explicit calculation of the state fidelity after error correction.

In QECCs that correct single-qubit errors, the error operators are
represented by $\sigma _l^\alpha $ with $l=1,2,\cdots ,L$ and $\alpha =1,2,3$
[11]. Let $R_0=I$, denoting the identity operator, and $R_{3\left(
l-1\right) +\alpha }=\sigma _l^\alpha $. We only consider orthogonal QECCs.
Most of the discovered QECCs belong to this class [5-12]. For orthogonal
QECCs, the encoded input state $\left| \Psi \left( 0\right) \right\rangle $
should satisfy the condition [10] 
\begin{equation}
\left\langle \Psi \left( 0\right) \right| R_n^{\dagger }R_{n^{^{\prime
}}}\left| \Psi \left( 0\right) \right\rangle =\delta _{nn^{^{\prime }}}\text{
}\left( n=0,1,\cdots ,3L\right) .  \label{17}
\end{equation}
During the error correction procedure, we first detect the error syndrome.
If there is a $R_n$ error, i.e., the state becomes $R_n\left| \Psi \left(
0\right) \right\rangle $, we apply the recovery operator $R_n^{-1}=R_n$ to
the state and thus get the correct initial state $\left| \Psi \left(
0\right) \right\rangle $. In the case of spatially correlated decoherence,
the error operators are represented by $Q_n$ $\left( n=0,1,\cdots ,3L\right) 
$, but we still adopt the above error correction procedure. If there is a $%
Q_n$ error, which occurs with probability $p_n$, we detect the error
syndrome and with probability $\left| \left\langle \Psi \left( 0\right)
\right| R_m^{\dagger }Q_n\left| \Psi \left( 0\right) \right\rangle \right|
^2 $ find that the error is $R_m$. After this detection, the state is
collapsed into $R_m\left| \Psi \left( 0\right) \right\rangle $. We apply the
recovery operator and thus get the initial state. The whole error correction
procedure described above yields the following average state fidelity after
error correction 
\begin{eqnarray}
F_a\left( \Delta t\right) &=&\stackrel{3L}{\stackunder{n=0}{\sum }}\stackrel{%
3L}{\stackunder{m=0}{\sum }}p_n\left| \left\langle \Psi \left( 0\right)
\right| R_m^{\dagger }Q_n\left| \Psi \left( 0\right) \right\rangle \right| ^2
\label{18} \\
&&  \nonumber \\
\ &=&1-o\left( \Delta t^2\right) .  \nonumber
\end{eqnarray}
In deriving Eq. (18), we have used Eqs. (11) and (17), together with the
identity $\stackunder{l,\alpha }{\sum }\left| U_{kl}^{\gamma \alpha }\right|
^2=1$ (from the unitarity of the matrix $U$ ). After the whole time $T_0$,
the final average state fidelity $F_a\left( T_0\right) $ is then
approximated by 
\begin{equation}
F_a\left( T_0\right) \simeq \left[ F_a\left( \Delta t\right) \right]
^N\simeq 1-o\left( N\Delta t^2\right) .  \label{19}
\end{equation}
Since $N\Delta t^2\propto \frac 1N$, it can be made arbitrarily small by a
frequent repetition of the error correction procedure. This demonstrates
that the QECCs devised to correct single-qubit errors are applicable for
reducing spatially correlated decoherence.

Before ending the paper, we should emphasize that we have omitted all the
other terms except the first one in the Hamiltonian (1). If these omitted
terms become important due to some special reason, the QECCs that correct
for single-qubit errors may not work well any more. For example, if the
second term in the Hamiltonian (1) is nor negligible, the quantum jump
errors will include not only the terms that can be expressed as sums of
single-qubit operators, but also the terms that are tensor products of them.
To combat this kind of decoherence, the QECCs need at least having the
ability to correct two-qubit errors.\\

\textbf{Acknowledgment}

This project was supported by the National Natural Science Foundation of
China.

\newpage\

\end{document}